\documentclass[twocolumn,showpacs,superscriptaddress,nofootinbib]{revtex4-1}
\usepackage{graphicx}
\usepackage{subfigure}
\usepackage{epsfig}
\usepackage{epstopdf}
\usepackage{color}
\usepackage{amsmath}

\newcommand{\be}{\begin{equation}}
\newcommand{\ee}{\end{equation}}
\newcommand{\ba}{\begin{eqnarray}}
\newcommand{\ea}{\end{eqnarray}}
\newcommand{\beq}{\begin{equation}}
\newcommand{\eeq}{\end{equation}}
\newcommand{\beqa}{\begin{eqnarray}}
\newcommand{\eeqa}{\end{eqnarray}}

\usepackage[normalem]{ulem}

\begin{document}

\title{Finely Split Phase Transitions of Rotating and Accelerating Black Holes}

\author{Niloofar Abbasvandi}
\email{niloofar.abbasvandi@uwaterloo.ca}
\address{Department of Physics and Astronomy, University of Waterloo, Waterloo,
Ontario, Canada, N2L 3G1}

\author{Wasif Ahmed}
\email{wasif.ahmed.bd@gmail.com}
\address{Department of Physics and Astronomy, University of Waterloo, Waterloo,
Ontario, Canada, N2L 3G1}
\address{Perimeter Institute, 31 Caroline St., Waterloo, Ontario, N2L 2Y5, Canada}

\author{Wan Cong}
\email{wcong@uwaterloo.ca}
\address{Department of Physics and Astronomy, University of Waterloo, Waterloo,
Ontario, Canada, N2L 3G1}
\address{Perimeter Institute, 31 Caroline St., Waterloo, Ontario, N2L 2Y5, Canada}

\author{David Kubiz\v n\'ak}
\email{dkubiznak@perimeterinstitute.ca}
\address{Perimeter Institute, 31 Caroline St., Waterloo, Ontario, N2L 2Y5, Canada}
\address{Department of Physics and Astronomy, University of Waterloo, Waterloo,
Ontario, Canada, N2L 3G1}

\author{Robert B. Mann}
\email{rbmann@uwaterloo.ca}
\address{Department of Physics and Astronomy, University of Waterloo, Waterloo,
Ontario, Canada, N2L 3G1}
\address{Perimeter Institute, 31 Caroline St., Waterloo, Ontario, N2L 2Y5, Canada}

\date{June 7, 2019}

\begin{abstract}
We investigate the thermodynamic phase behaviour of rotating and slowly accelerating AdS black holes.  While we find some similarities with the non-rotating charged counterparts, such as the
peculiar phenomena of `snapping swallow tails'  we also find subtle but significant distinctions  that can be attributed to a qualitatively modified parameter space of the solution. Consequently the zeroth order phase transition now occurs over a range of pressures, mini-entropic black holes no longer exist in the regime of slow acceleration, and the `no black hole region' emerges continuously---from a zero temperature extremal black hole.
 The formerly equal transition pressure now experiences a fine splitting, as the emergence of a no black hole region and the zeroth-order phase transition appear at different pressures, different also from the termination pressure of the first order phase transition.
This has the further effect
of admitting 
reentrant phase transitions that can be achieved in two ways, either
by varying the temperature at fixed pressure or varying the pressure at fixed temperature.
\end{abstract}

\maketitle

\section{Introduction}
\label{sec: intro}

Thermodynamics of black holes in AdS spacetimes can be studied in extended phase space \cite{Caldarelli:1999xj,Kastor2009, Dolan:2010ha, Cvetic:2010jb, Kubiznak:2016qmn} by treating the cosmological constant as thermodynamic pressure $P$. Together with its conjugate variable, the thermodynamic volume $V$, the effect of this is to add a $P-V$ term to the first law of black hole thermodynamics. The addition of this term calls for a comparison of the thermodynamic behaviour of black holes to  those of traditionally studied thermodynamic systems, such as regarding the efficiency of the corresponding heat engine \cite{Johnson:2014yja} and whether or not phase transitions are present. Of particular relevance are the first order phase transitions exhibited by charged AdS black holes in the canonical (fixed charged) ensemble. These are completely analogous to the familiar liquid-gas transitions observed in Van der Waals' gases \cite{Kubiznak:2012wp}, both being accompanied by the signature swallowtail-like curve in the free energy vs. temperature ($F-T$) graph \cite{Chamblin:1999tk, Kubiznak:2012wp}.
There are now many examples  of different families of black holes exhibiting a broad range
of   thermodynamic behaviour, so much so that the subject has come to be called Black Hole
Chemistry \cite{Kubiznak:2016qmn}.

Within this range of families, it is only recently that attention has turned toward accelerating
black holes. Described by the generalized C-metric \cite{Podolsky:2002nk,Hong:2004dm,Griffiths:2005qp}, these are accelerating black hole spacetimes that have conical deficits along their polar axes \cite{Podolsky:2002nk}. The conical deficits can be thought of as being caused by a cosmic string running through the spacetime, pulling on the black hole at its north and south poles with unbalanced tensions and causing its acceleration.  If the spacetime is asymptotically AdS and the acceleration is sufficiently small,  these spacetimes have only a single black hole horizon, and so it is possible to consider these systems in thermodynamic equilibrium. Numerous attempts at finding the corresponding thermodynamic variables, especially the black hole mass \cite{Appels:2016uha,Appels:2017xoe,Gregory:2017ogk,Astorino:2016ybm,Astorino:2016xiy}, were finally resolved in \cite{Anabalon:2018ydc,Anabalon:2018qfv}.   The final expression for the mass is consistent among various known approaches (including the conformal method, the Euclidean action calculation, and the holographic stress-energy calculation), and satisfies the corresponding first law of black hole thermodynamics and the associated Smarr formula.

In this paper we study the thermodynamic behaviour of accelerating and rotating black holes in AdS.
Our investigation is motivated by a recent study of the  phase transitions of charged and (slowly) accelerating black holes \cite{Abbasvandi:2018vsh}, in which the surprising phenomenon of ``{\em snapping swallowtails}'' was discovered:  the swallowtail indicating first order phase transitions snaps and disappears as one lowers the pressure  below a critical  transition pressure
\be\label{PtQ}
P_t^{(Q)}=\frac{3\mu^2}{8\pi Q^2}\,,
\ee
where $\mu$ is the differential tension causing the black hole to accelerate and $Q$ is its charge.
  This behaviour was attributed  to the presence of a ``point X'' in the space of parameters characterizing the black hole solution. Solutions near this point have peculiar characteristics. In particular, their isoperimetric ratio \cite{Cvetic:2010jb} becomes unbounded, earning them the designation \textit{mini-entropic}. The snapping swallowtail also heralds several new phenomena, namely (i) the emergence of a `no black hole' region in the $P-T$ plane, (ii) a termination of first order phase transitions and (iii) the phenomenon of zeroth order phase transition \cite{Altamirano:2013ane}, all happening at the single pressure at which the swallowtail snaps.
In addition, a reentrant phase transition was also observed to take place for some values of the string tension.  The characteristics of the `snapping point' were further analyzed recently in \cite{Gregory:2019dtq}.

In this paper we show that all of the above phenomena observed for the charged case remain present for the rotating case but with a number of subtle and interesting differences.
Namely, unlike the charged case, `point X' no longer belongs to the admissible parameter  space and consequently  slowly accelerating rotating uncharged mini-entropic black holes do not exist.
Although the swallow tail still  snaps at \cite{Gregory:2019dtq}
\be\label{PtJ}
P_t^{(J)}=\frac{3(1-2\mu)^2}{8\pi |J|}\frac{x_0\sqrt{2x_0+1}}{2}\,,
\ee
where $J$ is the angular momentum of the black hole,
and
\be
x_0=\frac{\sqrt{1+12C^2}-1}{3}\,,\quad C=\frac{\mu}{1-2\mu}\,,
\ee
this happens for black holes outside of the slowly accelerating regime. Consequently, instead of one transition pressure $P_t^{(J)}$ as in \cite{Abbasvandi:2018vsh}, we now observe 3 important critical pressures:  the pressure $P_{nbh}$ at which the `no black hole' region appears,  the pressure $P_0<P_{nbh}$
where the zeroth order phase transition starts, and the pressure $P_f<P_0$ at which zeroth   and first order phase transitions co-terminate.  In this sense the presence of rotation provides a ``fine splitting'' of the transition pressure $P_t^{(J)}$.

\section{Thermodynamics of Rotating, Accelerating Black Holes}
\label{sec: TD}
 The accelerating and rotating black hole is described by the following generalized C-metric \cite{Podolsky:2002nk,Hong:2004dm,Griffiths:2005qp}:
\ba\label{eq: metric}
ds^2 &=& \frac{1}{\Omega^2}\bigg\{ -\frac{f}{\Sigma}\bigg[\frac{dt}{\alpha}-a \sin^2\theta\frac{d\phi}{K}\bigg]^2 + \frac{\Sigma}{f}dr^2\nonumber\\ & & + \frac{\Sigma \, r^2}{g}d\theta^2+\frac{g\, \sin^2\!\theta}{\Sigma\, r^2}\bigg[\frac{a\, dt}{\alpha}-(r^2+a^2)\frac{d\phi}{K}\bigg]^2\bigg\}\,, \qquad
\ea
where
\ba
f&=&(1-A^2r^2)\Big(1-\frac{2m}{r}+\frac{a^2}{r^2}\Big)+\frac{r^2+a^2}{l^2}\,,\nonumber\\
g&=&1+2mA\cos\theta+\Big( a^2A^2-\frac{a^2}{l^2}\Big)\cos^2\!\theta\,,\nonumber\\
\Sigma&=&1+\frac{a^2}{r^2}\cos^2\!\theta\,, \quad \alpha = \sqrt{1-A^2l^2}\,,\nonumber\\
 \Omega &=& 1+Ar\cos\theta\,.
\ea
The solution is parameterized by 5 parameters: the rotation parameter $a$, the mass parameter $m$, the cosmological length-scale $l$, the acceleration parameter $A$, and the (overall) conical deficit $K$. These are loosely related to 5 physical charges:
 the angular momentum $J$, the mass $M$, the cosmological pressure
\be
P=-\frac{\Lambda}{8\pi}=\frac{3}{8\pi l^2}\,,
\ee
and the north pole and south pole conical deficits, corresponding to the following string tensions:\footnote{ That 
string tensions (or conical defects) emerging from the black hole should be regarded as true hair, i.e. a new charge that the black hole can carry, was recently elucidated in  \cite{Gregory:2019dtq}.}
\be
\mu_{\pm}=\frac{1}{4}\bigg[1-\frac{\Xi\pm2mA}{K}\bigg] = \frac{1}{4}\bigg[1-\frac{K_{\pm}}{K}\bigg]\,,
 \ee
 where
\be
\Xi = 1-\frac{a^2}{l^2}+a^2A^2\,.
\ee
The parameter $\alpha$ is not independent and simply normalizes the timelike Killing vector $\partial_t$ \cite{Anabalon:2018qfv,Anabalon:2018ydc}. The angle coordinate $\phi$ has periodicity $2\pi$. The conformal boundary of the spacetime is located at $\Omega=0$.  We refer to \cite{Podolsky:2003gm} for further analysis of this metric, including the Penrose diagrams.

\begin{figure*}
    \centering
    \subfigure{\includegraphics[scale = 0.97]{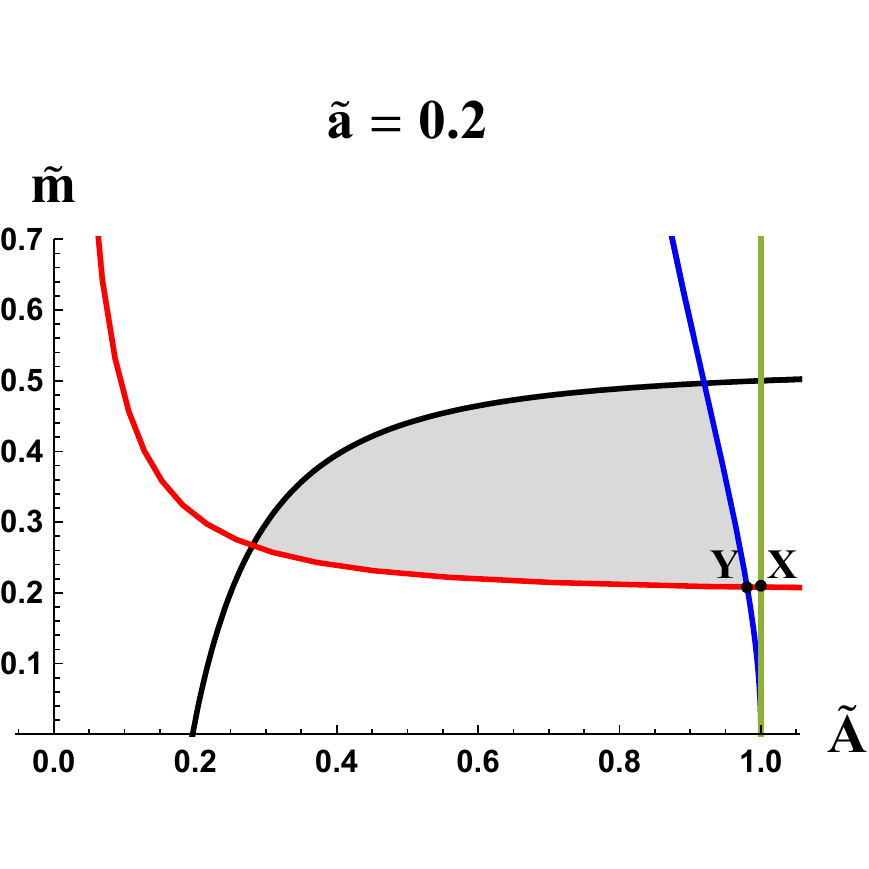}} \quad \quad
    \subfigure{\includegraphics[scale = 0.97]{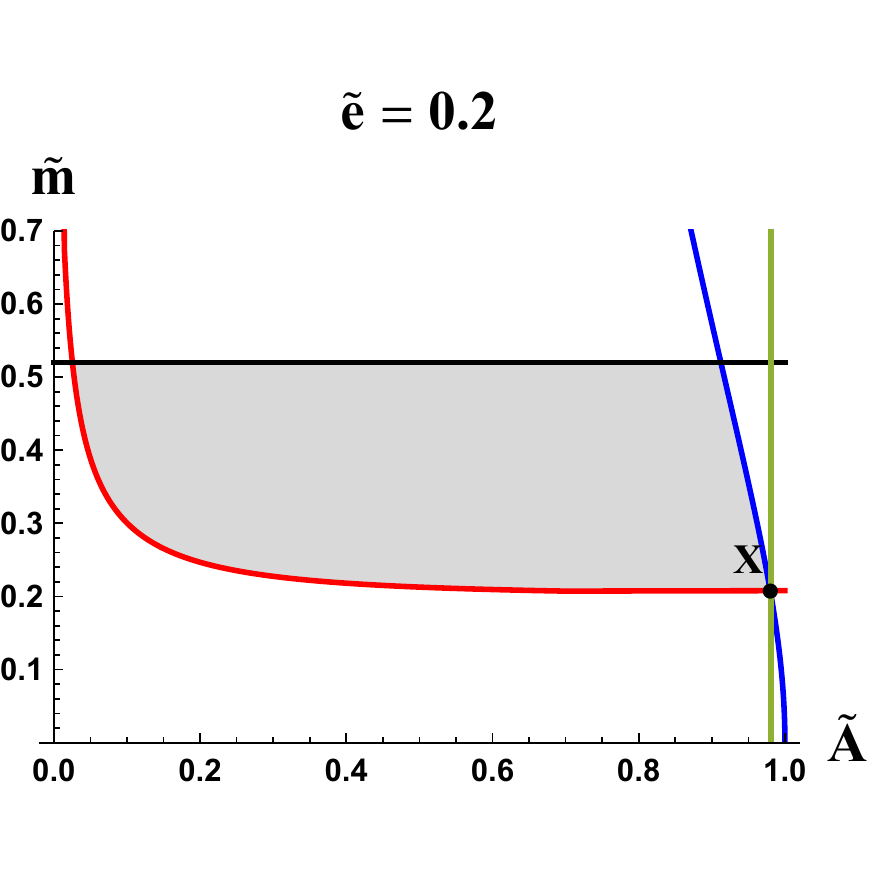}}
    \caption{{\bf Admissible parameter space comparison. } The admissible parameter space for rotating (left) and charged (right) slowly accelerating black holes is displayed by the shaded region in the $\tilde A-\tilde m$ plane. The left figure, displayed for fixed $\tilde a=0.2$,  summarizes the constraints discussed in the main text as follows: the black, red, and blue curves respectively correspond to the boundaries of the Lorentzian metric signature, the existence of a bulk black hole, and the absence of an acceleration horizon; the green line displays $\alpha = 0$. The right figure (reproduced from \cite{Abbasvandi:2018vsh}) displays the analogous curves for the charged case with a fixed dimensionless charge parameter $\tilde e = e A = 0.2$.
    While in the charged case the point $X$ lies at the intersection of three (red, blue, and green) boundary curves, rotation removes this degeneracy and we now observe two salient intersection points: $X$ and $Y$.}
    \label{fig: rotating PS}
\end{figure*}

For solutions with small acceleration $A$, the only horizons present in the spacetime are the black hole horizons. As the acceleration increases, additional horizons such as acceleration and cosmological horizons can develop. The former regime is known as the {\em slowly accelerating regime}.
From the thermodynamic point of view this regime is especially interesting  as the spacetime contains only one black hole and can be assigned a unique (black hole horizon) temperature. The
associated thermodynamic variables are
 \cite{Anabalon:2018ydc,Anabalon:2018qfv}:
\ba
M&=& \frac{m(\Xi+a^2/l^2)(1-A^2 l^2\Xi)}{K\Xi\alpha(1+a^2A^2)}\,,\\
T&=& \frac{f'_+\, r^2_+}{4\pi\alpha(r^2_++a^2)}\,, \quad
S=\frac{\pi (r^2_++a^2)}{K(1-A^2r_+^2)}\,,
\\J&=& \frac{ma}{K^2}\,,\,\,\Omega = \frac{aK}{\alpha(r^2_++a^2)}+\frac{aK(1-A^2l^2\Xi)}{l^2\Xi\alpha(1+a^2A^2)}\,,
\\V&=&\frac{4\pi}{3K\alpha} \Big[ \frac{r_+(r_+^2+a^2)}{(1-A^2 r_+^2)^2}\nonumber
\\&&+ \frac{m[a^2(1-A^2l^2\Xi)+A^2l^4\Xi(\Xi+a^2/l^2)]}{(1+a^2A^2)\Xi}\Big]\, ,\\
\lambda_\pm &=&\frac{m}{\alpha}\frac{\big[\Xi+\frac{a^2}{l^2}(2-A^2l^2\Xi)\big]}{(1+a^2A^2)\Xi^2}-\frac{r_+}{\alpha(1\pm Ar_+)}\nonumber\\
& &\pm\frac{A l^2 (\Xi+a^2/l^2)}{\alpha(1+a^2A^2)}\,.
\ea
Here $V$ is the thermodynamic volume, a conjugate quantity to the thermodynamic pressure \cite{Kubiznak:2016qmn}, and $\lambda_\pm$ are the thermodynamic lengths, conjugates to the string tensions \cite{Appels:2017xoe}. The above quantities satisfy both the generalized first law and the associated Smarr relation:
\ba
\delta M&=&T\delta S+\Omega \delta J+\lambda_+\delta \mu_+
+\lambda_-\delta \mu_- +V\delta P\,,\nonumber\\
M&=&2(TS+J\Omega-PV)\,.
\ea
Note that, due to their dimensionless character, the string tension terms do not modify the Smarr formula.

As discussed above, the parameter $K$ is a physical parameter---it corresponds to a choice of the distribution of the conical deficits (or cosmic strings in the physical picture). Perhaps the most physically intuitive is the situation where only one cosmic string is present and the black hole is suspended on it---without loss of generality we choose the north axis to be regular. This corresponds to setting
$K=K_+$, so that
\be
\label{eq: mu}
\mu_+=0\,,\quad \mu\equiv\mu_- = \frac{mA}{K_+}.
\ee
For the charged case, the presence of the second string has only quantitative but not qualitative effects on the phase behaviour of the system  \cite{Abbasvandi:2018vsh}  and that is why we choose
\eqref{eq: mu}  for the rotating case as well.

\section{Parameter Space}
\label{sec: PS}
 Before turning to the phase transitions, we first probe
the admissible parameter space for which slowly accelerating rotating black holes exist and compare it to the parameter space of the charged accelerating black holes analyzed in \cite{Abbasvandi:2018vsh}.

As we have already fixed the parameter $K$ by imposing \eqref{eq: mu}, our family of black holes is characterized by the remaining 4 parameters $\{a,A,l,m\}$.
To simplify the analysis let us introduce the dimensionless parameters
\be
\label{eq: dimless}
\tilde m=mA\,,\quad \tilde a=aA\,,\quad \tilde A=Al\,.
\ee
These are constrained by the requirements of (i) having the right metric signature, (ii) existence of black holes in the bulk, and (iii) the absence of acceleration and cosmological horizons (for the validity of equilibrium thermodynamics).

These constraints dictate the admissible parameter space in the dimensionless $(\tilde A,\tilde m,\tilde a)$ plane. Two dimensional slices of this three-dimensional parameter space can easily be displayed;  see Fig.~\ref{fig: rotating PS} a for a choice of $\tilde a=0.2$; the full admissible parameter region would be a union over $\tilde a$ of such slices.

The first of the above conditions requires $g>0$, yielding the constraint
\be \tilde m < \frac{1}{2}\Xi=\frac{1}{2}+ \frac{1}{2}\bigg(\tilde a^2 -\frac{\tilde a^2}{\tilde A^2}\bigg)\,,\ee
the boundaries of which yield the black curve in the left diagram in Fig. \ref{fig: rotating PS}.

Second, for a black hole to exist within the boundaries of the spacetime, we require that the function $f$ has at least two roots counted with multiplicity in the range $r\in(0,1/A)$. A double root corresponds to the presence of an extremal black hole.
The condition for a double root and hence an extremal black hole is
\be
f(r_+) = 0 = f'(r_+)\,,
\ee
and yields complicated equations for $\tilde A = \tilde A(\tilde r_+,\tilde a)$ and $\tilde m = \tilde m(\tilde r_+,\tilde a)$. These can be plotted parametrically, using $\tilde r_+=r_+/l$ as the parameter, and are displayed in the left diagram in Fig. \ref{fig: rotating PS} as a red curve. Solutions on this curve represent extremal black holes while those above this curve represent non-extremal bulk black holes.

The last condition requires that there are no additional horizons in the bulk besides the black hole horizon. An additional horizon in the bulk first appears when $f$ develops a root on the boundary ($\Omega=0$), which in terms of new coordinates
\begin{equation}
    x = \frac{1}{Ar}\,,\quad y = \cos\theta\,,
\end{equation}
simply reads $x=-y$. We thus have to solve a condition
for the existence of an extremal horizon on the boundary, which is $f(x=-y)=0=f'(x=-y)$.\footnote{Of course, this is the same as looking for a double root of $f(r)$ in the range $r\in(-1/A,0)$ and $r>1/A$, which corresponds to an extremal horizon in the bulk that `goes all the way to the boundary'.} This yields the following equations:
\ba
\tilde m &=& \frac{y(1+\tilde a^2 y^2)^2}{1-y^2(\tilde a^2+3)-\tilde a^2 y^4}\,,\nonumber\\
\tilde A &=& \frac{\sqrt{1-y^2(\tilde a^2+3)-\tilde a^2 y^4}}{(1-y^2)\sqrt{1-\tilde a^2y^2}}\,,
\ea
which again can be plotted parameterically, with $y\in[-1,1]$ now playing the role of a parameter. The result  is displayed in the left diagram in   Fig.~\ref{fig: rotating PS} by the blue curve. The spacetimes with additional (acceleration and cosmological) horizons are to the right of this curve and are excluded by the slow acceleration condition.

Putting everything together, it is now obvious that the parameter space of the rotating black holes (Fig.~\ref{fig: rotating PS}a) is qualitatively different from the parameter space of the charged black holes  (Fig.~\ref{fig: rotating PS}b).
While in the charged case the point $X$ lies at the intersection of
three (red, blue, and green) boundary curves, rotation removes this degeneracy and we now observe
two salient intersection points: $X$ and $Y$.
The point $X$, at the intersection of green and red curves, corresponds to $P^{(J)}_t$ given by \eqref{PtJ}, or
\be\label{Ptpert}
P^{(J)}_t\approx \frac{3\mu^2}{8\pi |J|}\Big(1-\mu^2+O(\mu^4)\Bigr)\,.
\ee
It lies strictly outside the admissible region. This prevents the formation of mini-entropic black holes \cite{Abbasvandi:2018vsh}, which exist in the vicinity of the point $X$ displayed in Fig. \ref{fig: rotating PS}, and whose thermodynamic volume grow unbounded whilst their entropies remain finite.

\begin{figure}[t]
    \centering
    \includegraphics[width=0.98\linewidth, height=0.32\textheight]{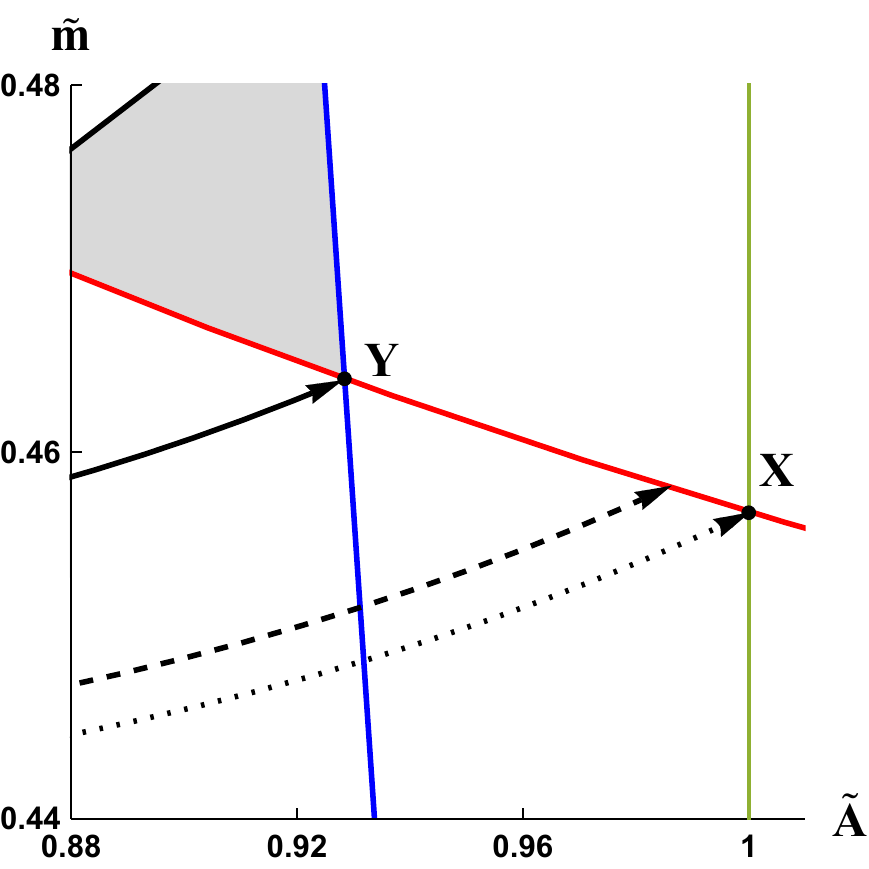}
    \caption{\textbf{Swallowtail slicing of the parameter space.}
     The arrows display the projections of three important swallowtails (with $J=1$ and $\mu=0.24$) into a single $\tilde a=0.4$ plane of the parameter space near the points $X$ and $Y$.
     The solid black arrow corresponds to $P=P_{nbh}\approx0.00688$ and terminates at the point $Y$.
     On the $F-T$ diagram, this corresponds to a curve whose extremal black hole ($T=0$ point) lies just within the admissible parameter space. Note that the arrow \textit{does} lie within the admissible parameter space for all $\tilde a\in [0,0.4]$. This is not obvious in the figure because the admissible region shrinks with increasing $\tilde a$. The dashed arrow displays the swallowtail with $P = P_0\approx 0.0062$. Unlike the previous curve, it does not lie fully within the parameter space (and rather terminates at the slow acceleration boundary). It corresponds to a swallowtail similar to the one shown on Fig \ref{fig: ZOT}, but with the red point lying directly below the free-energy peak of the swallowtail. The dotted arrow is for $P = P_t \approx 0.00601$ and reaches the point $X$. Like the dashed case, it lies partly outside of the admissible region. Strictly speaking, swallowtails with different pressures terminate at different $\tilde a$ planes. However, for points $X$ and $Y$ the variation of $\tilde a$ is of the order $O(\mu^5)$ and can be in our figure neglected. }
    \label{fig: rotating PS2}
\end{figure}

The point $Y$, located at the intersection of blue and red curves, marks the emergence of the no black hole region and occurs at $P_{nbh}$.
At this pressure, the metric function develops two double roots---corresponding to the extremal black hole in the bulk and an extremal horizon at the boundary, giving 
\be
\tilde m=\tilde a+\tilde a^3\,,\quad \tilde A=\frac{1}{\sqrt{1+\tilde a^2}}\,.
\ee
Equating these with \eqref{swallowtails} below yields
\be\label{Pnbh}
P_{nbh}=\frac{3\mu^2}{8\pi |J|}\,.
\ee

For small $\mu$, this is very close to the transition pressure \eqref{PtJ}, as is clear from
\eqref{Ptpert}. Note also that two important pressures $P_0$ and $P_f$  necessarily lie in between these two pressures; we shall discuss these in the next section.

An additional piece of information is encoded in how the  $\mu,\, J\,, P=$ const. curves cut through the parameter space. Such curves correspond to the `swallow-tail curves' of the free energy in the  $F-T$ diagrams. To display these, we use the following 3 equations:
\be
J=\frac{3 \tilde a \tilde m}{8\pi P \tilde A^2 K_+^2}\,,\quad \mu=\frac{\tilde m}{K_+}\,,\quad K_+=1-\frac{\tilde a^2}{\tilde A^2}+\tilde a^2+2\tilde m\,,
\ee
to find
\ba\label{swallowtails}
\tilde m&=&\frac{(1+\tilde a^2)\mu^2}{8\pi \tilde a  PJ/3+\mu-2\mu^2}\,,\nonumber\\
\tilde A&=&\frac{\sqrt{\tilde a^2 +3\mu \tilde a(1-2\mu)/(8\pi PJ)}}{\sqrt{1+\tilde a^2}}\,,
\ea
which can be plotted parameterically over $\tilde a$. For illustration, the projections of the three important swallow tails at $P_{nbh}, P_0\approx P_f$, and $P_t^{(J)}$ for $J=1$ and $\mu=0.24$ are displayed in Fig.~\ref{fig: rotating PS2}.
The fact that the latter two terminate at the slow acceleration boundary has  far reaching consequences for the phase transitions studied in the next section.

\section{Phase Transitions}
\label{sec: PT}
\begin{figure}
    \centering
    \includegraphics[width=0.98\linewidth, height=0.32\textheight]{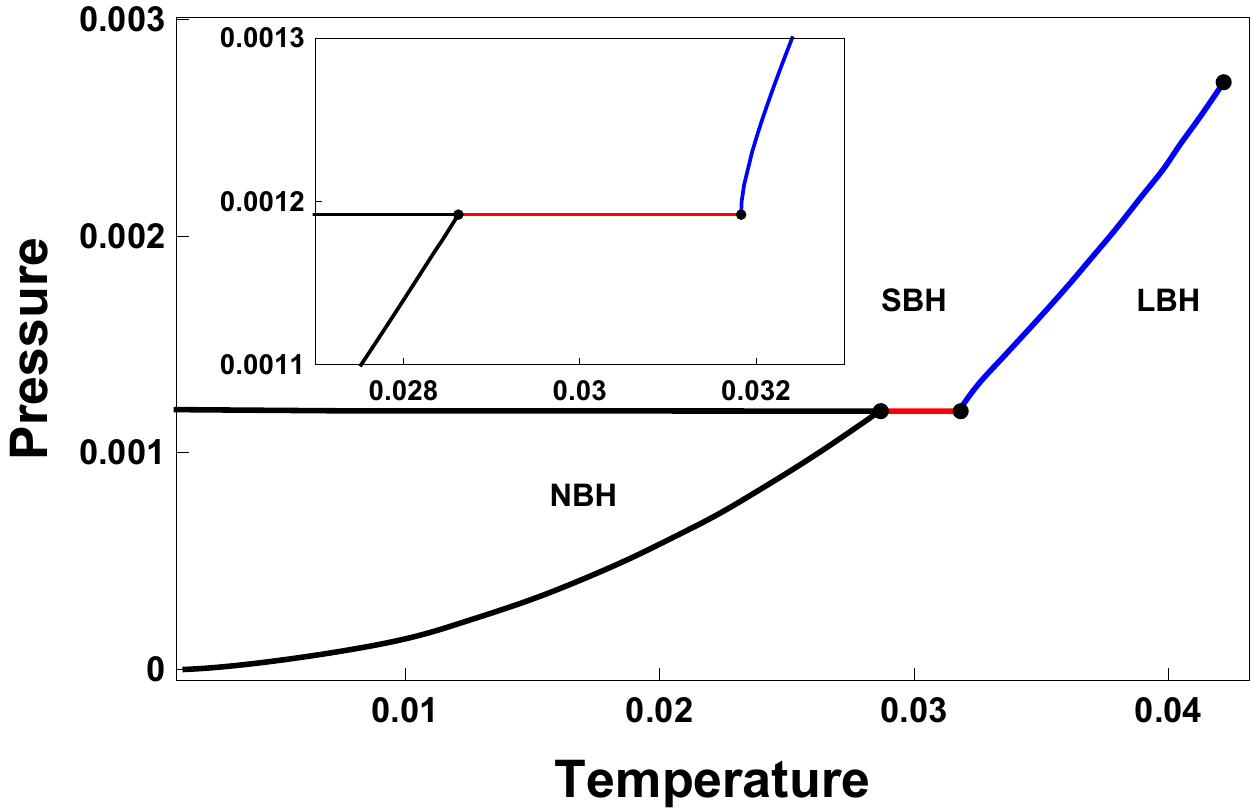}
    \caption{{\bf Phase diagram: $\mu = 0.1$.} For small tensions, the $P-T$ phase diagram is similar to the charged case, with termination of the first order coexistence line (blue), the presence of a zeroth order transition line (red) and a `no black hole' (NBH) region delimited by the black lines. The first order transition separates small black hole (SBH) and large black hole (LBH) solutions. The inset shows an enlarged view near $P = P_f$, where the first order phase transition terminates.
    At this $\mu$,  we have $P_0\approx P_f\approx P_{nbh}\approx0.001194$. }
    \label{fig: PT}
    \end{figure}

In this section we study the thermodynamic behaviour and possible phase transitions of our family of rotating accelerating black holes.
We show that, although similar in many aspects, the behaviour is richer than in the charged case \cite{Abbasvandi:2018vsh}. Perhaps most intriguing is the phenomenon of `fine splitting' of the transition pressure discussed below.

As always, the thermodynamic equilibrium corresponds to the global minimum of the free energy
\be
F=M-TS=F(T,P,J,\mu)\,.
\ee
In our considerations we set $J=1$ and plot the corresponding $P-T$ phase diagrams for various string tensions $\mu$, see Figs.~\ref{fig: PT} and ~\ref{fig: PT2}.

\subsection{Fine Splitting of Transition Pressures}

For small tensions, the phase diagram is qualitatively similar to that of the charged accelerating black hole, compare Fig.~\ref{fig: PT} with Fig.~7a in \cite{Abbasvandi:2018vsh}. Namely, we observe a first order phase transition (displayed by the blue curve),  which terminates at a critical point at $P=P_c$ on one end and at a termination point at $P=P_f$ on the other end. (Whereas the critical point at $P_c$ is a standard feature of many black holes, the existence of the termination point seems unique for accelerating black holes.) From this termination point a zeroth-order phase transition coexistence line (displayed by the red curve) emerges and eventually ends at $P=P_0$ on the boundary (black curve) of the no black hole region, which appears at $P=P_{nbh}$.

\begin{figure}[h]
    \centering
    \includegraphics[width=0.98\linewidth, height=0.32\textheight]
    {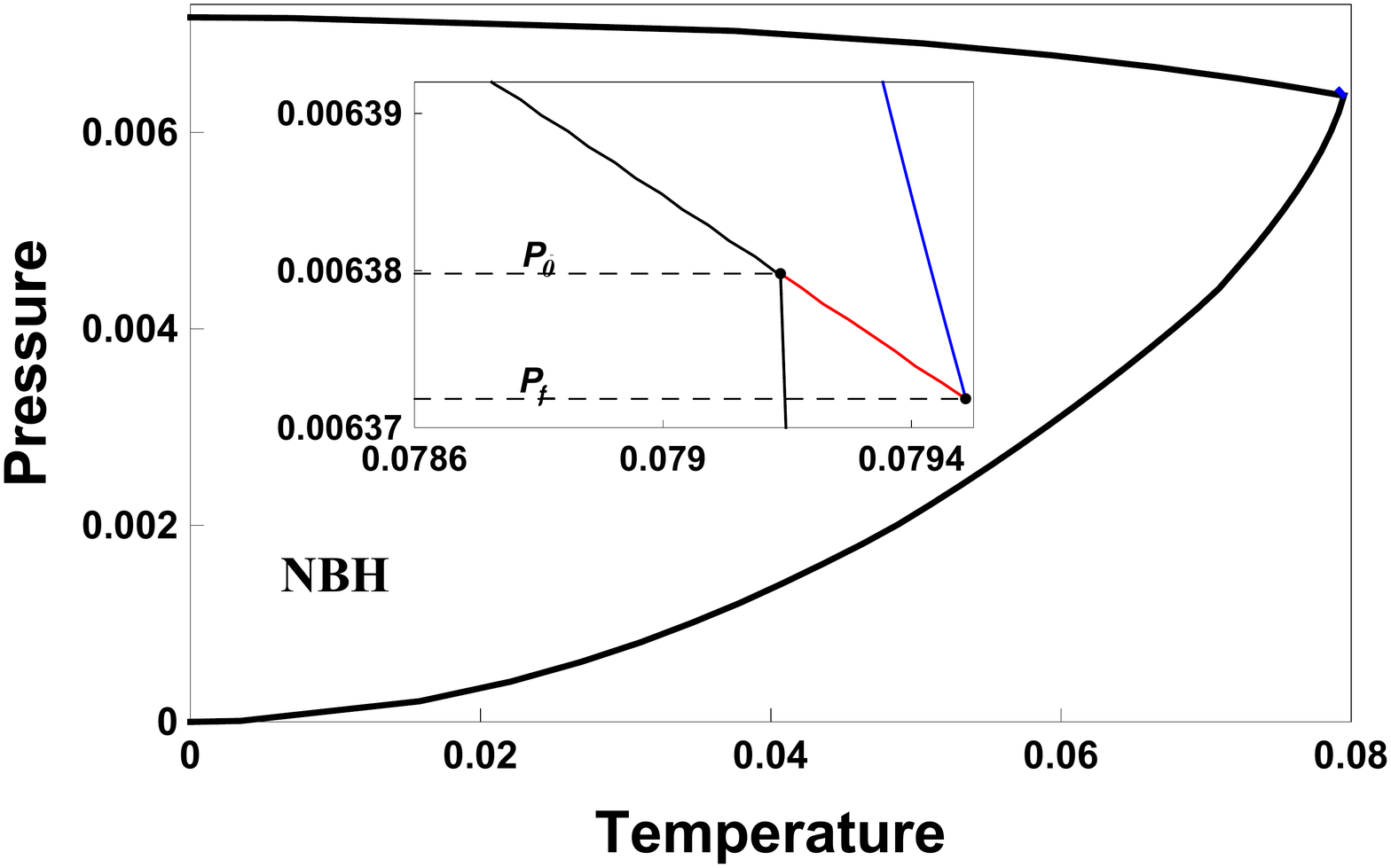}
    \caption{{\bf Phase diagram: $\mu = 0.245$.} For this $\mu$, the $P-T$ diagram clearly illustrates the fine splitting of the transition pressures. Comparing with the $\mu = 0.1$ case, the first order phase transition and zeroth order phase transition curves shrank considerably. However, the inset shows that the zeroth order transition line has a non-trivial negative gradient, and the gradient of the first order phase transition went from positive to negative. For this tension we have  $P_{nbh}\approx 0.007165, P_{0}\approx 0.00637981, P_f\approx 0.00637184$, and $P_t^{(J)}\approx0.006215$, all apparently distinct.}
    \label{fig: PT2}
\end{figure}

In the charged case these pressures are in fact exactly equal, given by the transition pressure $P_t^{(Q)}$, \eqref{PtQ}.
However, in the rotating case this is only true approximately and the situation is in fact much more subtle. Namely, these pressures order as follows:
\be
P_{nbh}\gtrsim P_0\gtrsim P_f\gtrsim P_t^{(J)}\,,
\ee
and though they seem indistinguishable in Fig.~\ref{fig: PT}, the distinction is
 `more apparent' for bigger string tensions $\mu$, as shown in Fig.~\ref{fig: PT2} and its inset.

\begin{figure}
    \centering
    \includegraphics[width=0.98\linewidth, height=0.28\textheight]{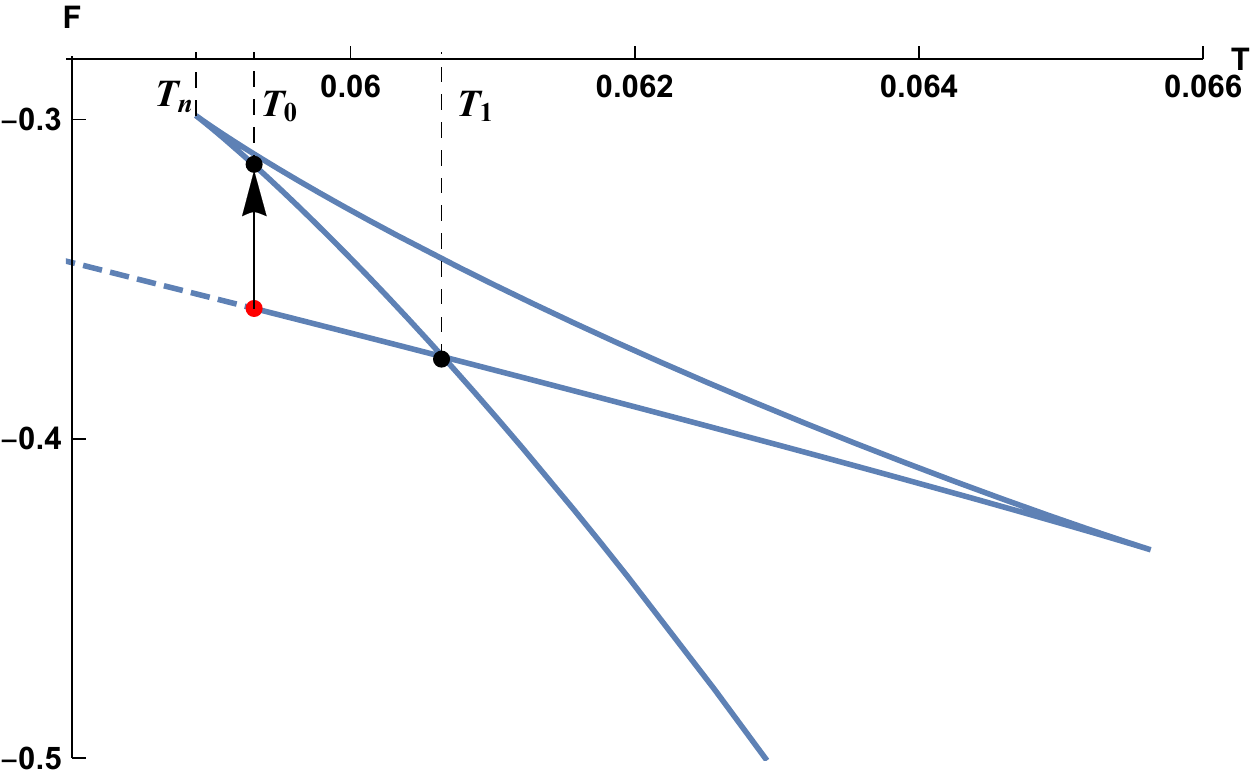}
    \caption{{\bf Slow acceleration bound and its effect on phase transitions. } The $F-T$ diagram shows a swallowtail at a pressure $P_f<P<P_0$. The `last' slowly accelerating black hole is highlighted by a red dot -- the part of the lower temperature branch of black holes (that develop extra horizons) denoted by a dashed curve is removed. This is the origin of the temperature driven reentrant phase transition as well as of the no black hole region. At any temperature, the solution with the lowest free energy is thermodynamically preferred.  Starting from high temperatures, these are solutions lying on the large black hole branch of the swallowtail. As the temperature decreases, we move up the swallowtail, with the black hole radii decreasing continuously in the process. At $T = T_1$, a first order phase transition occurs as the thermodynamically favoured state moves from the large black hole branch to the small black hole branch, with a discontinuity in the black hole radius. As the temperature decreases further to $T_0$, a zeroth order phase transition occurs, when the small black hole transits into an intermediate black hole with a jump in the free-energy. At this pressure, the no black hole region starts at $T = T_n$, the temperature of the upper cusp.}
    \label{fig: ZOT}
\end{figure}

The origin of this fine splitting can be traced to the qualitatively different parameter space in the rotating case, and in particular to the different behaviour of the slow acceleration bound. In the charged case, such a bound does not play any role in equilibrium thermodynamics: it is completely absent for $P>P_t^{(Q)}$, and below $P_t^{(Q)}$ it only removes small unstable black holes in the upper branch of the free energy that do not correspond to the global minimum. In the rotating case, however, this bound can occur in the stable branch and thus affects the thermodynamic behaviour, as illustrated in Fig.~\ref{fig: ZOT}.

The appearance of a no black hole region, at $P_{nbh}$, corresponds to the pressure at which the last slowly accelerating black hole (displayed by a red dot in Fig.~\ref{fig: ZOT}) coincides with an extremal black hole on the lower branch of the swallowtail.  As pressure decreases, the last slowly accelerating black hole `moves to the right', eliminating a branch of black holes to the left (denoted by the dashed curve) that no longer obey the slow acceleration condition. As a consequence, the no black hole region continuously grows larger, until a zeroth order phase transition  appears at $P_{0}$, where the black hole at the red dot has the same temperature as the upper cusp of the swallow tail. For a slightly smaller pressure, there develops a small range of temperatures for which the lower branch is already removed but there is a corresponding new branch of black holes (on the lower left branch of the upper left cusp in Fig.~\ref{fig: ZOT}) that become stable, and we observe a zeroth-order phase transition. This situation holds until the pressure $P_{f}$ is reached for which the red dot black hole moves all the way to the intersection of the swallow tail, at which point both the zeroth--order and the first order phase transitions simultaneously disappear. For even smaller pressures, the red dot black hole keeps moving further right and no longer occurs in the stable branch. This behaviour finally terminates at the transition pressure $P_t^{(J)}$ for which the swallow tail snaps and completely disappears, as in the case of a charged black hole \cite{Abbasvandi:2018vsh}. 

In other words, in the rotating case the interesting phase behaviour happens above the pressure for which the the actual snapping occurs. This is a consequence of the existence of solutions on the small black hole branch of the swallowtail that develop acceleration horizons and thus do not lie in the admissible parameter space;  the corresponding analysis is depicted in Fig.~\ref{fig: rotating PS2}.

\subsection{Reentrant phase transitions}

\begin{figure}[h]
    \centering
    \includegraphics[scale = 0.65]{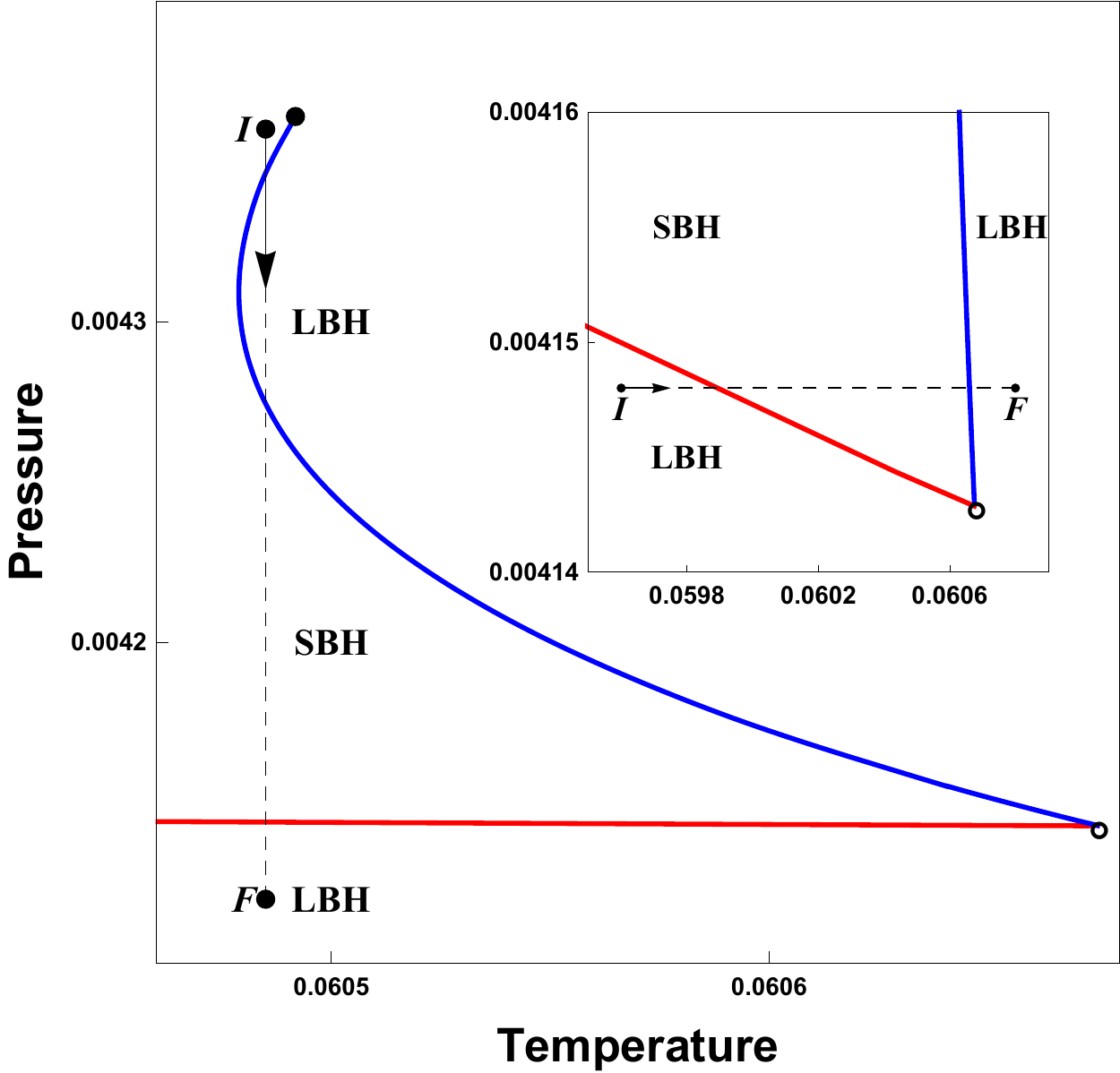}
    \caption{ {\bf Pressure and temperature driven reentrant phase transitions.} First order (blue)
    and zeroth order (red)  coexistence lines are displayed for $\mu=0.19$. For this tension the first order line is double-valued, which together with the zeroth-order phase transition, gives rise to a pressure driven SBH/LBH/SBH/LBH reentrant phase transition as we move from the initial point $I$ to the final point $F$ in the diagram.  The inset dislays the temperature driven LBH/SBH/LBH phase transition, innate to the rotating case, that happens in the tiny range of pressure $P\in (P_f,P_0)$, as we move from the initial point $I$ to the final point $F$ in the inset.}    \label{fig: RPT2}
\end{figure}

The fact that $P_0>P_f$ results in a non-trivial negative slope of the zeroth-order coexistence line. As a consequence there will be a temperature driven reentrant phase transition for $P_0>P>P_f$. For any fixed $P$ in this range, there is a large to small to large black hole phase transition as the temperature monotonically increases, as shown in the inset of Fig.~\ref{fig: RPT2}.

This behaviour can be seen from a different perspective in  Fig. \ref{fig: ZOT}. Beginning at high temperatures $T>T_f$ the stable solution is on the lowest branch of the swallowtail, corresponding to a large black hole.  As the temperature decreases
the solution moves up and to the left of the swallowtail  with the black hole radius decreasing continuously. At $T = T_f$ a first order phase transition occurs as the thermodynamically favoured state moves from the large black hole branch to the small black hole branch at the lower left,
with a discontinuity in the black hole radius. As  $T$ decreases further, once $T=T_0$ a zeroth order phase transition occurs, when the small black hole transits into an intermediate (large) black hole with a jump in the free-energy.  The corresponding branch finally disappears  at $T=T_n$, which marks the onset of the no black hole region -- as no (slowly accelerating) black holes exist below $T_n$.

As tension increases, the slope of the first order coexistence line becomes negative in a certain range of temperatures and we can observe a presence of another, this time pressure driven, reentrant phase transition. Particularly interesting is a tiny range of tensions around $\mu\approx 0.19$ for which
the first order phase transition coexistence line becomes double valued (its gradient changes from positive to negative as the pressure decreases) and for a fixed pressure can be `crossed twice',
 as shown in main part of Fig.~\ref{fig: RPT2}.\footnote{For large values of $\mu$ the gradient of the first order phase transition coexistence line is everywhere negative.}
This results in a complicated reentrant SBH/LBH/SBH/LBH phase transition as the pressure monotonically decreases from $I$ to $F$, crossing the first order coexistence line twice and zeroth-order coexistence line once.
Let us stress that whereas the similar pressure driven reentrant phase transition has been already observed for the charged case \cite{Abbasvandi:2018vsh}, the temperature driven reentrant phase transition described above is new.

\section{Conclusions}

We have discovered that the thermodynamic behaviour of rotating accelerating black holes has subtle but significant distinctions as compared to the their charged counterparts.  The key distinction between them is that the coexistence line in a $P-T$ plot separating small black holes from large ones and the NBH region induces a `fine splitting' of transition pressures.  Namely, we observe
a sequence of pressures
$P_{nbh} > P_0 > P_f$
in which the minimum pressure possible for a small black hole
as $T\to 0$ is larger than the pressure $P_0$ at the onset of a zeroth order phase transition, which is larger still than the pressure $P_f$ at which the zeroth and first order phase transitions merge. Although the swallow tail still snaps at $P_t^{(J)}<P_f$, this does not affect the global minimum of the free energy and thence has no effect on the equilibrium thermodynamics.
This is in contrast to the charged case, for which $P_{nbh} = P_0 = P_f=P_t^{(J)}$, and which in turn leads to the existence of mini-entropic black holes  \cite{Abbasvandi:2018vsh}.
 The splitting is fine in
that $P_{nbh}\approx  P_0 \approx P_f\approx P_t^{(J)}$, with the correction of the order of $\mu^2$, c.f. Eqs. \eqref{Ptpert} and \eqref{Pnbh}.  Likewise, the lifting of this pressure degeneracy  means that mini-entropic black holes do not exist in the rotating case.

Furthermore, this splitting has an interesting physical consequence in that it admits two kinds of reentrant phase transitions,   one at fixed $T$ where the pressure monotonically decreases and one at fixed $P$ in the range  $P_0 > P > P_f$
where the temperature monotonically increases. Both these types of phase transitions (although qualitatively different from our case) are also present for black holes with triple points \cite{Altamirano:2013uqa}.

It is clear that accelerating black holes contain interesting but often subtle new thermodynamic phenomena, and there are still things to be learned.  While a full study including both charge and rotation remains to be carried out,  of particular interest would be to go beyond the slow-acceleration regime.  This would involve the presence of two horizons, with all the associated difficulties this scenario entails \cite{Kubiznak:2015bya}.   However it might be possible to circumvent these problems by
either by placing the system in a cavity \cite{Simovic:2018tdy,Simovic:2019zgb} or by adjusting parameters to recover thermodynamic equilibrium \cite{Mbarek:2018bau}, approaches that have proved to be successful for asymptotically de Sitter black holes.

\section*{Acknowledgments}
\label{sc:acknowledgements}

This work was supported in part by the Natural Sciences and Engineering Research Council of Canada.
D.K. acknowledges the Perimeter Institute for Theoretical Physics  for support. Research at Perimeter Institute is supported by the Government of Canada through the Department of Innovation, Science and Economic Development Canada and by the Province of Ontario through the Ministry of Research, Innovation and Science.


\begin{thebibliography}{27}%
\makeatletter
\providecommand \@ifxundefined [1]{%
 \@ifx{#1\undefined}
}%
\providecommand \@ifnum [1]{%
 \ifnum #1\expandafter \@firstoftwo
 \else \expandafter \@secondoftwo
 \fi
}%
\providecommand \@ifx [1]{%
 \ifx #1\expandafter \@firstoftwo
 \else \expandafter \@secondoftwo
 \fi
}%
\providecommand \natexlab [1]{#1}%
\providecommand \enquote  [1]{``#1''}%
\providecommand \bibnamefont  [1]{#1}%
\providecommand \bibfnamefont [1]{#1}%
\providecommand \citenamefont [1]{#1}%
\providecommand \href@noop [0]{\@secondoftwo}%
\providecommand \href [0]{\begingroup \@sanitize@url \@href}%
\providecommand \@href[1]{\@@startlink{#1}\@@href}%
\providecommand \@@href[1]{\endgroup#1\@@endlink}%
\providecommand \@sanitize@url [0]{\catcode `\\12\catcode `\$12\catcode
  `\&12\catcode `\#12\catcode `\^12\catcode `\_12\catcode `\%12\relax}%
\providecommand \@@startlink[1]{}%
\providecommand \@@endlink[0]{}%
\providecommand \url  [0]{\begingroup\@sanitize@url \@url }%
\providecommand \@url [1]{\endgroup\@href {#1}{\urlprefix }}%
\providecommand \urlprefix  [0]{URL }%
\providecommand \Eprint [0]{\href }%
\providecommand \doibase [0]{http://dx.doi.org/}%
\providecommand \selectlanguage [0]{\@gobble}%
\providecommand \bibinfo  [0]{\@secondoftwo}%
\providecommand \bibfield  [0]{\@secondoftwo}%
\providecommand \translation [1]{[#1]}%
\providecommand \BibitemOpen [0]{}%
\providecommand \bibitemStop [0]{}%
\providecommand \bibitemNoStop [0]{.\EOS\space}%
\providecommand \EOS [0]{\spacefactor3000\relax}%
\providecommand \BibitemShut  [1]{\csname bibitem#1\endcsname}%
\let\auto@bib@innerbib\@empty
\bibitem [{\citenamefont {Caldarelli}\ \emph {et~al.}(2000)\citenamefont
  {Caldarelli}, \citenamefont {Cognola},\ and\ \citenamefont
  {Klemm}}]{Caldarelli:1999xj}%
  \BibitemOpen
  \bibfield  {author} {\bibinfo {author} {\bibfnamefont {M.~M.}\ \bibnamefont
  {Caldarelli}}, \bibinfo {author} {\bibfnamefont {G.}~\bibnamefont {Cognola}},
  \ and\ \bibinfo {author} {\bibfnamefont {D.}~\bibnamefont {Klemm}},\ }\href
  {\doibase 10.1088/0264-9381/17/2/310} {\bibfield  {journal} {\bibinfo
  {journal} {Class. Quant. Grav.}\ }\textbf {\bibinfo {volume} {17}},\ \bibinfo
  {pages} {399} (\bibinfo {year} {2000})},\ \Eprint
  {http://arxiv.org/abs/hep-th/9908022} {arXiv:hep-th/9908022 [hep-th]}
  \BibitemShut {NoStop}%
\bibitem [{\citenamefont {Kastor}\ \emph {et~al.}(2009)\citenamefont {Kastor},
  \citenamefont {Ray},\ and\ \citenamefont {Traschen}}]{Kastor2009}%
  \BibitemOpen
  \bibfield  {author} {\bibinfo {author} {\bibfnamefont {D.}~\bibnamefont
  {Kastor}}, \bibinfo {author} {\bibfnamefont {S.}~\bibnamefont {Ray}}, \ and\
  \bibinfo {author} {\bibfnamefont {J.}~\bibnamefont {Traschen}},\ }\href
  {http://stacks.iop.org/0264-9381/26/i=19/a=195011} {\bibfield  {journal}
  {\bibinfo  {journal} {Classical and Quantum Gravity}\ }\textbf {\bibinfo
  {volume} {26}},\ \bibinfo {pages} {195011} (\bibinfo {year}
  {2009})}\BibitemShut {NoStop}%
\bibitem [{\citenamefont {Dolan}(2011)}]{Dolan:2010ha}%
  \BibitemOpen
  \bibfield  {author} {\bibinfo {author} {\bibfnamefont {B.~P.}\ \bibnamefont
  {Dolan}},\ }\href {\doibase 10.1088/0264-9381/28/12/125020} {\bibfield
  {journal} {\bibinfo  {journal} {Class. Quant. Grav.}\ }\textbf {\bibinfo
  {volume} {28}},\ \bibinfo {pages} {125020} (\bibinfo {year} {2011})},\
  \Eprint {http://arxiv.org/abs/1008.5023} {arXiv:1008.5023 [gr-qc]}
  \BibitemShut {NoStop}%
\bibitem [{\citenamefont {Cvetic}\ \emph {et~al.}(2011)\citenamefont {Cvetic},
  \citenamefont {Gibbons}, \citenamefont {Kubiznak},\ and\ \citenamefont
  {Pope}}]{Cvetic:2010jb}%
  \BibitemOpen
  \bibfield  {author} {\bibinfo {author} {\bibfnamefont {M.}~\bibnamefont
  {Cvetic}}, \bibinfo {author} {\bibfnamefont {G.~W.}\ \bibnamefont {Gibbons}},
  \bibinfo {author} {\bibfnamefont {D.}~\bibnamefont {Kubiznak}}, \ and\
  \bibinfo {author} {\bibfnamefont {C.~N.}\ \bibnamefont {Pope}},\ }\href
  {\doibase 10.1103/PhysRevD.84.024037} {\bibfield  {journal} {\bibinfo
  {journal} {Phys. Rev.}\ }\textbf {\bibinfo {volume} {D84}},\ \bibinfo {pages}
  {024037} (\bibinfo {year} {2011})},\ \Eprint {http://arxiv.org/abs/1012.2888}
  {arXiv:1012.2888 [hep-th]} \BibitemShut {NoStop}%
\bibitem [{\citenamefont {Kubiznak}\ \emph {et~al.}(2017)\citenamefont
  {Kubiznak}, \citenamefont {Mann},\ and\ \citenamefont
  {Teo}}]{Kubiznak:2016qmn}%
  \BibitemOpen
  \bibfield  {author} {\bibinfo {author} {\bibfnamefont {D.}~\bibnamefont
  {Kubiznak}}, \bibinfo {author} {\bibfnamefont {R.~B.}\ \bibnamefont {Mann}},
  \ and\ \bibinfo {author} {\bibfnamefont {M.}~\bibnamefont {Teo}},\ }\href
  {http://stacks.iop.org/0264-9381/34/i=6/a=063001} {\bibfield  {journal}
  {\bibinfo  {journal} {Classical and Quantum Gravity}\ }\textbf {\bibinfo
  {volume} {34}},\ \bibinfo {pages} {063001} (\bibinfo {year}
  {2017})}\BibitemShut {NoStop}%
\bibitem [{\citenamefont {Johnson}(2014)}]{Johnson:2014yja}%
  \BibitemOpen
  \bibfield  {author} {\bibinfo {author} {\bibfnamefont {C.~V.}\ \bibnamefont
  {Johnson}},\ }\href {\doibase 10.1088/0264-9381/31/20/205002} {\bibfield
  {journal} {\bibinfo  {journal} {Class. Quant. Grav.}\ }\textbf {\bibinfo
  {volume} {31}},\ \bibinfo {pages} {205002} (\bibinfo {year} {2014})},\
  \Eprint {http://arxiv.org/abs/1404.5982} {arXiv:1404.5982 [hep-th]}
  \BibitemShut {NoStop}%
\bibitem [{\citenamefont {Kubiznak}\ and\ \citenamefont
  {Mann}(2012)}]{Kubiznak:2012wp}%
  \BibitemOpen
  \bibfield  {author} {\bibinfo {author} {\bibfnamefont {D.}~\bibnamefont
  {Kubiznak}}\ and\ \bibinfo {author} {\bibfnamefont {R.~B.}\ \bibnamefont
  {Mann}},\ }\href {\doibase 10.1007/JHEP07(2012)033} {\bibfield  {journal}
  {\bibinfo  {journal} {JHEP}\ }\textbf {\bibinfo {volume} {07}},\ \bibinfo
  {pages} {033} (\bibinfo {year} {2012})},\ \Eprint
  {http://arxiv.org/abs/1205.0559} {arXiv:1205.0559 [hep-th]} \BibitemShut
  {NoStop}%
\bibitem [{\citenamefont {Chamblin}\ \emph {et~al.}(1999)\citenamefont
  {Chamblin}, \citenamefont {Emparan}, \citenamefont {Johnson},\ and\
  \citenamefont {Myers}}]{Chamblin:1999tk}%
  \BibitemOpen
  \bibfield  {author} {\bibinfo {author} {\bibfnamefont {A.}~\bibnamefont
  {Chamblin}}, \bibinfo {author} {\bibfnamefont {R.}~\bibnamefont {Emparan}},
  \bibinfo {author} {\bibfnamefont {C.~V.}\ \bibnamefont {Johnson}}, \ and\
  \bibinfo {author} {\bibfnamefont {R.~C.}\ \bibnamefont {Myers}},\ }\href
  {\doibase 10.1103/PhysRevD.60.064018} {\bibfield  {journal} {\bibinfo
  {journal} {Phys. Rev.}\ }\textbf {\bibinfo {volume} {D60}},\ \bibinfo {pages}
  {064018} (\bibinfo {year} {1999})},\ \Eprint
  {http://arxiv.org/abs/hep-th/9902170} {arXiv:hep-th/9902170 [hep-th]}
  \BibitemShut {NoStop}%
\bibitem [{\citenamefont {Podolsky}(2002)}]{Podolsky:2002nk}%
  \BibitemOpen
  \bibfield  {author} {\bibinfo {author} {\bibfnamefont {J.}~\bibnamefont
  {Podolsky}},\ }\href {\doibase 10.1023/A:1013961411430} {\bibfield  {journal}
  {\bibinfo  {journal} {Czech. J. Phys.}\ }\textbf {\bibinfo {volume} {52}},\
  \bibinfo {pages} {1} (\bibinfo {year} {2002})},\ \Eprint
  {http://arxiv.org/abs/gr-qc/0202033} {arXiv:gr-qc/0202033 [gr-qc]}
  \BibitemShut {NoStop}%
\bibitem [{\citenamefont {Hong}\ and\ \citenamefont {Teo}(2005)}]{Hong:2004dm}%
  \BibitemOpen
  \bibfield  {author} {\bibinfo {author} {\bibfnamefont {K.}~\bibnamefont
  {Hong}}\ and\ \bibinfo {author} {\bibfnamefont {E.}~\bibnamefont {Teo}},\
  }\href {\doibase 10.1088/0264-9381/22/1/007} {\bibfield  {journal} {\bibinfo
  {journal} {Class. Quant. Grav.}\ }\textbf {\bibinfo {volume} {22}},\ \bibinfo
  {pages} {109} (\bibinfo {year} {2005})},\ \Eprint
  {http://arxiv.org/abs/gr-qc/0410002} {arXiv:gr-qc/0410002 [gr-qc]}
  \BibitemShut {NoStop}%
\bibitem [{\citenamefont {Griffiths}\ and\ \citenamefont
  {Podolsky}(2006)}]{Griffiths:2005qp}%
  \BibitemOpen
  \bibfield  {author} {\bibinfo {author} {\bibfnamefont {J.~B.}\ \bibnamefont
  {Griffiths}}\ and\ \bibinfo {author} {\bibfnamefont {J.}~\bibnamefont
  {Podolsky}},\ }\href {\doibase 10.1142/S0218271806007742} {\bibfield
  {journal} {\bibinfo  {journal} {Int. J. Mod. Phys.}\ }\textbf {\bibinfo
  {volume} {D15}},\ \bibinfo {pages} {335} (\bibinfo {year} {2006})},\ \Eprint
  {http://arxiv.org/abs/gr-qc/0511091} {arXiv:gr-qc/0511091 [gr-qc]}
  \BibitemShut {NoStop}%
\bibitem [{\citenamefont {Appels}\ \emph {et~al.}(2016)\citenamefont {Appels},
  \citenamefont {Gregory},\ and\ \citenamefont {Kubiznak}}]{Appels:2016uha}%
  \BibitemOpen
  \bibfield  {author} {\bibinfo {author} {\bibfnamefont {M.}~\bibnamefont
  {Appels}}, \bibinfo {author} {\bibfnamefont {R.}~\bibnamefont {Gregory}}, \
  and\ \bibinfo {author} {\bibfnamefont {D.}~\bibnamefont {Kubiznak}},\ }\href
  {\doibase 10.1103/PhysRevLett.117.131303} {\bibfield  {journal} {\bibinfo
  {journal} {Phys. Rev. Lett.}\ }\textbf {\bibinfo {volume} {117}},\ \bibinfo
  {pages} {131303} (\bibinfo {year} {2016})},\ \Eprint
  {http://arxiv.org/abs/1604.08812} {arXiv:1604.08812 [hep-th]} \BibitemShut
  {NoStop}%
\bibitem [{\citenamefont {Appels}\ \emph {et~al.}(2017)\citenamefont {Appels},
  \citenamefont {Gregory},\ and\ \citenamefont {Kubiznak}}]{Appels:2017xoe}%
  \BibitemOpen
  \bibfield  {author} {\bibinfo {author} {\bibfnamefont {M.}~\bibnamefont
  {Appels}}, \bibinfo {author} {\bibfnamefont {R.}~\bibnamefont {Gregory}}, \
  and\ \bibinfo {author} {\bibfnamefont {D.}~\bibnamefont {Kubiznak}},\ }\href
  {\doibase 10.1007/JHEP05(2017)116} {\bibfield  {journal} {\bibinfo  {journal}
  {JHEP}\ }\textbf {\bibinfo {volume} {05}},\ \bibinfo {pages} {116} (\bibinfo
  {year} {2017})},\ \Eprint {http://arxiv.org/abs/1702.00490} {arXiv:1702.00490
  [hep-th]} \BibitemShut {NoStop}%
\bibitem [{\citenamefont {Gregory}(2017)}]{Gregory:2017ogk}%
  \BibitemOpen
  \bibfield  {author} {\bibinfo {author} {\bibfnamefont {R.}~\bibnamefont
  {Gregory}},\ }\bibfield  {booktitle} {\emph {\bibinfo {booktitle}
  {{Proceedings, 3rd Karl Schwarzschild Meeting on Gravitational Physics and
  the Gauge/Gravity Correspondence (KSM 2017): Frankfurt am Main, Germany, July
  24-28, 2017}}},\ }\href {\doibase 10.1088/1742-6596/942/1/012002} {\bibfield
  {journal} {\bibinfo  {journal} {J. Phys. Conf. Ser.}\ }\textbf {\bibinfo
  {volume} {942}},\ \bibinfo {pages} {012002} (\bibinfo {year} {2017})},\
  \Eprint {http://arxiv.org/abs/1712.04992} {arXiv:1712.04992 [hep-th]}
  \BibitemShut {NoStop}%
\bibitem [{\citenamefont {Astorino}(2017)}]{Astorino:2016ybm}%
  \BibitemOpen
  \bibfield  {author} {\bibinfo {author} {\bibfnamefont {M.}~\bibnamefont
  {Astorino}},\ }\href {\doibase 10.1103/PhysRevD.95.064007} {\bibfield
  {journal} {\bibinfo  {journal} {Phys. Rev.}\ }\textbf {\bibinfo {volume}
  {D95}},\ \bibinfo {pages} {064007} (\bibinfo {year} {2017})},\ \Eprint
  {http://arxiv.org/abs/1612.04387} {arXiv:1612.04387 [gr-qc]} \BibitemShut
  {NoStop}%
\bibitem [{\citenamefont {Astorino}(2016)}]{Astorino:2016xiy}%
  \BibitemOpen
  \bibfield  {author} {\bibinfo {author} {\bibfnamefont {M.}~\bibnamefont
  {Astorino}},\ }\href {\doibase 10.1016/j.physletb.2016.07.019} {\bibfield
  {journal} {\bibinfo  {journal} {Phys. Lett.}\ }\textbf {\bibinfo {volume}
  {B760}},\ \bibinfo {pages} {393} (\bibinfo {year} {2016})},\ \Eprint
  {http://arxiv.org/abs/1605.06131} {arXiv:1605.06131 [hep-th]} \BibitemShut
  {NoStop}%
\bibitem [{\citenamefont {Anabalon}\ \emph
  {et~al.}(2018{\natexlab{a}})\citenamefont {Anabalon}, \citenamefont {Appels},
  \citenamefont {Gregory}, \citenamefont {Kubiznak}, \citenamefont {Mann},\
  and\ \citenamefont {Ovgun}}]{Anabalon:2018ydc}%
  \BibitemOpen
  \bibfield  {author} {\bibinfo {author} {\bibfnamefont {A.}~\bibnamefont
  {Anabalon}}, \bibinfo {author} {\bibfnamefont {M.}~\bibnamefont {Appels}},
  \bibinfo {author} {\bibfnamefont {R.}~\bibnamefont {Gregory}}, \bibinfo
  {author} {\bibfnamefont {D.}~\bibnamefont {Kubiznak}}, \bibinfo {author}
  {\bibfnamefont {R.~B.}\ \bibnamefont {Mann}}, \ and\ \bibinfo {author}
  {\bibfnamefont {A.}~\bibnamefont {Ovgun}},\ }\href@noop {} {\  (\bibinfo
  {year} {2018}{\natexlab{a}})},\ \Eprint {http://arxiv.org/abs/1805.02687}
  {arXiv:1805.02687 [hep-th]} \BibitemShut {NoStop}%
\bibitem [{\citenamefont {Anabalon}\ \emph
  {et~al.}(2018{\natexlab{b}})\citenamefont {Anabalon}, \citenamefont {Gray},
  \citenamefont {Gregory}, \citenamefont {Kubiznak},\ and\ \citenamefont
  {Mann}}]{Anabalon:2018qfv}%
  \BibitemOpen
  \bibfield  {author} {\bibinfo {author} {\bibfnamefont {A.}~\bibnamefont
  {Anabalon}}, \bibinfo {author} {\bibfnamefont {F.}~\bibnamefont {Gray}},
  \bibinfo {author} {\bibfnamefont {R.}~\bibnamefont {Gregory}}, \bibinfo
  {author} {\bibfnamefont {D.}~\bibnamefont {Kubiznak}}, \ and\ \bibinfo
  {author} {\bibfnamefont {R.~B.}\ \bibnamefont {Mann}},\ }\href@noop {} {\
  (\bibinfo {year} {2018}{\natexlab{b}})},\ \Eprint
  {http://arxiv.org/abs/1811.04936} {arXiv:1811.04936 [hep-th]} \BibitemShut
  {NoStop}%
\bibitem [{\citenamefont {Abbasvandi}\ \emph {et~al.}(2018)\citenamefont
  {Abbasvandi}, \citenamefont {Cong}, \citenamefont {Kubiznak},\ and\
  \citenamefont {Mann}}]{Abbasvandi:2018vsh}%
  \BibitemOpen
  \bibfield  {author} {\bibinfo {author} {\bibfnamefont {N.}~\bibnamefont
  {Abbasvandi}}, \bibinfo {author} {\bibfnamefont {W.}~\bibnamefont {Cong}},
  \bibinfo {author} {\bibfnamefont {D.}~\bibnamefont {Kubiznak}}, \ and\
  \bibinfo {author} {\bibfnamefont {R.~B.}\ \bibnamefont {Mann}},\ }\href@noop
  {} {\  (\bibinfo {year} {2018})},\ \Eprint {http://arxiv.org/abs/1812.00384}
  {arXiv:1812.00384 [gr-qc]} \BibitemShut {NoStop}%
\bibitem [{\citenamefont {Altamirano}\ \emph {et~al.}(2013)\citenamefont
  {Altamirano}, \citenamefont {Kubiznak},\ and\ \citenamefont
  {Mann}}]{Altamirano:2013ane}%
  \BibitemOpen
  \bibfield  {author} {\bibinfo {author} {\bibfnamefont {N.}~\bibnamefont
  {Altamirano}}, \bibinfo {author} {\bibfnamefont {D.}~\bibnamefont
  {Kubiznak}}, \ and\ \bibinfo {author} {\bibfnamefont {R.~B.}\ \bibnamefont
  {Mann}},\ }\href {\doibase 10.1103/PhysRevD.88.101502} {\bibfield  {journal}
  {\bibinfo  {journal} {Phys. Rev.}\ }\textbf {\bibinfo {volume} {D88}},\
  \bibinfo {pages} {101502} (\bibinfo {year} {2013})},\ \Eprint
  {http://arxiv.org/abs/1306.5756} {arXiv:1306.5756 [hep-th]} \BibitemShut
  {NoStop}%
\bibitem [{\citenamefont {Gregory}\ and\ \citenamefont
  {Scoins}(2019)}]{Gregory:2019dtq}%
  \BibitemOpen
  \bibfield  {author} {\bibinfo {author} {\bibfnamefont {R.}~\bibnamefont
  {Gregory}}\ and\ \bibinfo {author} {\bibfnamefont {A.}~\bibnamefont
  {Scoins}},\ }\href@noop {} {\  (\bibinfo {year} {2019})},\ \Eprint
  {http://arxiv.org/abs/1904.09660} {arXiv:1904.09660 [hep-th]} \BibitemShut
  {NoStop}%
\bibitem [{\citenamefont {Podolsky}\ \emph {et~al.}(2003)\citenamefont
  {Podolsky}, \citenamefont {Ortaggio},\ and\ \citenamefont
  {Krtous}}]{Podolsky:2003gm}%
  \BibitemOpen
  \bibfield  {author} {\bibinfo {author} {\bibfnamefont {J.}~\bibnamefont
  {Podolsky}}, \bibinfo {author} {\bibfnamefont {M.}~\bibnamefont {Ortaggio}},
  \ and\ \bibinfo {author} {\bibfnamefont {P.}~\bibnamefont {Krtous}},\ }\href
  {\doibase 10.1103/PhysRevD.68.124004} {\bibfield  {journal} {\bibinfo
  {journal} {Phys. Rev.}\ }\textbf {\bibinfo {volume} {D68}},\ \bibinfo {pages}
  {124004} (\bibinfo {year} {2003})},\ \Eprint
  {http://arxiv.org/abs/gr-qc/0307108} {arXiv:gr-qc/0307108 [gr-qc]}
  \BibitemShut {NoStop}%
\bibitem [{\citenamefont {Altamirano}\ \emph {et~al.}(2014)\citenamefont
  {Altamirano}, \citenamefont {Kubiznak}, \citenamefont {Mann},\ and\
  \citenamefont {Sherkatghanad}}]{Altamirano:2013uqa}%
  \BibitemOpen
  \bibfield  {author} {\bibinfo {author} {\bibfnamefont {N.}~\bibnamefont
  {Altamirano}}, \bibinfo {author} {\bibfnamefont {D.}~\bibnamefont
  {Kubiznak}}, \bibinfo {author} {\bibfnamefont {R.~B.}\ \bibnamefont {Mann}},
  \ and\ \bibinfo {author} {\bibfnamefont {Z.}~\bibnamefont {Sherkatghanad}},\
  }\href {\doibase 10.1088/0264-9381/31/4/042001} {\bibfield  {journal}
  {\bibinfo  {journal} {Class. Quant. Grav.}\ }\textbf {\bibinfo {volume}
  {31}},\ \bibinfo {pages} {042001} (\bibinfo {year} {2014})},\ \Eprint
  {http://arxiv.org/abs/1308.2672} {arXiv:1308.2672 [hep-th]} \BibitemShut
  {NoStop}%
\bibitem [{\citenamefont {Kubiznak}\ and\ \citenamefont
  {Simovic}(2016)}]{Kubiznak:2015bya}%
  \BibitemOpen
  \bibfield  {author} {\bibinfo {author} {\bibfnamefont {D.}~\bibnamefont
  {Kubiznak}}\ and\ \bibinfo {author} {\bibfnamefont {F.}~\bibnamefont
  {Simovic}},\ }\href {\doibase 10.1088/0264-9381/33/24/245001} {\bibfield
  {journal} {\bibinfo  {journal} {Class. Quant. Grav.}\ }\textbf {\bibinfo
  {volume} {33}},\ \bibinfo {pages} {245001} (\bibinfo {year} {2016})},\
  \Eprint {http://arxiv.org/abs/1507.08630} {arXiv:1507.08630 [hep-th]}
  \BibitemShut {NoStop}%
\bibitem [{\citenamefont {Simovic}\ and\ \citenamefont
  {Mann}(2019{\natexlab{a}})}]{Simovic:2018tdy}%
  \BibitemOpen
  \bibfield  {author} {\bibinfo {author} {\bibfnamefont {F.}~\bibnamefont
  {Simovic}}\ and\ \bibinfo {author} {\bibfnamefont {R.}~\bibnamefont {Mann}},\
  }\href {\doibase 10.1088/1361-6382/aaf445} {\bibfield  {journal} {\bibinfo
  {journal} {Class. Quant. Grav.}\ }\textbf {\bibinfo {volume} {36}},\ \bibinfo
  {pages} {014002} (\bibinfo {year} {2019}{\natexlab{a}})},\ \Eprint
  {http://arxiv.org/abs/1807.11875} {arXiv:1807.11875 [gr-qc]} \BibitemShut
  {NoStop}%
\bibitem [{\citenamefont {Simovic}\ and\ \citenamefont
  {Mann}(2019{\natexlab{b}})}]{Simovic:2019zgb}%
  \BibitemOpen
  \bibfield  {author} {\bibinfo {author} {\bibfnamefont {F.}~\bibnamefont
  {Simovic}}\ and\ \bibinfo {author} {\bibfnamefont {R.~B.}\ \bibnamefont
  {Mann}},\ }\href@noop {} {\bibfield  {journal} {\bibinfo  {journal}
  {Submitted to: J. High Energy Phys.}\ } (\bibinfo {year}
  {2019}{\natexlab{b}})},\ \Eprint {http://arxiv.org/abs/1904.04871}
  {arXiv:1904.04871 [gr-qc]} \BibitemShut {NoStop}%
\bibitem [{\citenamefont {Mbarek}\ and\ \citenamefont
  {Mann}(2019)}]{Mbarek:2018bau}%
  \BibitemOpen
  \bibfield  {author} {\bibinfo {author} {\bibfnamefont {S.}~\bibnamefont
  {Mbarek}}\ and\ \bibinfo {author} {\bibfnamefont {R.~B.}\ \bibnamefont
  {Mann}},\ }\href {\doibase 10.1007/JHEP02(2019)103} {\bibfield  {journal}
  {\bibinfo  {journal} {JHEP}\ }\textbf {\bibinfo {volume} {02}},\ \bibinfo
  {pages} {103} (\bibinfo {year} {2019})},\ \Eprint
  {http://arxiv.org/abs/1808.03349} {arXiv:1808.03349 [hep-th]} \BibitemShut
  {NoStop}%
\end{thebibliography}

%

\end{document}